\documentstyle[11pt,newpasp,twoside,epsf]{article}
\def\edcomment#1{\iffalse\marginpar{\raggedright\sl#1\/}\else\relax\fi}
\reversemarginpar

\begin{document}
\title{Early Hard X-ray Afterglows of Short GRBs with Konus Experiments}
\author{D.D.Frederiks, R.L.Aptekar, S.V.Golenetskii, V.N.Il'inskii, E.P.Mazets, and V.D.Palshin}
\affil{A.F.Ioffe Physico-Technical Institute, Politekhnicheskaya 26, St.Petersburg, 194021, Russia\\
e-mail: fred@mail.ioffe.ru}
\author{T.L.Cline}
\affil{NASA Goddard Space Flight Center, Code 661, Greenbelt, MD 20771, USA}

\begin{abstract}
For a ten of 125 short GRBs observed by Konus-Wind and the Konus-A the existance of 
statistically significant flux of hard photons accompanying initial event
for a time of tens to hundred seconds after the trigger was revealed. 
Temporal, spectral, and energetic characteristics of these events are presented.
The statistical analysis of the whole burst sample reveals that the afterglow is a more common 
feature of short GRBs.
\end{abstract}

\section{Introduction}
The gamma-ray burst (GRB) duration has a bimodal distribution (Mazets et al. (1981), Norris
et al. (1984); Hurley (1992); Kouveliotou et al. (1993)). This indicates the existence of two distinct
morphological classes of events, namely short-duration ($<$ 2 s) bursts and long-duration ($>$ 2 s)
bursts. Approximately 20 per cent of the observed bursts are short. Their energy spectra are usually
harder than the spectra of long bursts (Kouveliotou et al. 1993).

Searches for the optical and radio afterglow of short GRBs have been carried out in only a few
cases. Four GRBs were localized with high accuracy by the Interplanetary Network (IPN) (Hurley
et al. 2002). Nighther optical nor radio afterglow emission was detected for these four events. No X-ray
counterparts have been detected so far for the short bursts localized by Beppo-SAX (Gandolfi et al.
2000) or by HETE-2 (Lamb et al. 2002). The rapid follow-up observations have resulted in only upper
limits on the brightness of the afterglows from these GRBs.

At the same time early X-ray and gamma-ray afterglows of short bursts were detected in a bumber
cases by the GRB detectors themselves, in time intervals from seconds to tens of seconds after the
trigger. BATSE observations showed that such a weak afterglow exists for some of the short bursts,
lasting several tens seconds (Burenin, 2000; Lazatti et al. 2001; Connaughton, 2002). The afterglows
of short GRBs were also detected by the Konus-Wind experiment. Post--burst emission in the energy
range below 1 MeV is seen for about 10 per cent of events. 
The statistical analysis of the whole Konus--Wind and Konus--A short burst sample reveals that 
the afterglow is a more common feature of short GRBs. 

\section{Individual Bursts}

The early afterglows of short GRBs are observed for 11 short bursts, i.e. 
GRB 9511014a, 980605, 980706a, 981107, 990313, 990327, 990516, 990712a, 000218,
000701b, and 000727. These afterglows appear in time intervals from several seconds 
up to 100 seconds after the trigger.

A limited volume of this proceeding doesn't allow us to present all illustrations.
One can find lightcurves and spectra of these events as well as fluence, peak flux, 
and spectral parameters tables in the "Konus Catalog of Short GRBs" (Mazets et al. 2002).  
The electronic version of the catalog is available at http://www.ioffe.ru/LEA/shortGRBs/Catalog/.
Let us look now at a typical example.

Figure 1 presents on the left pannel light curves of GRB990712a (UT 27915.510 s) in three
energy windows: 10-50, 50-200, and 200-750 keV. On the right pannel there is the light curve
of postburst emission for combined 10-750 keV energy range. An emission is seen up to 150 seconds
after the burst trigger on the level up to $2 \times 10^{-7}$ erg cm$^{-2}$s$^{-1}$ compared to 
$5 \times 10^{-5}$ erg cm$^{-2}$s$^{-1}$ for the main event.

Most of the considered events are typical short GRBs with hard spectrum. On the left pannel of Figure 2 
an energetic spectrum of GRB990712a is shown. Solid line is a best fit by COMP model with $\alpha=-0.2$
and $E_{0}=610$keV. The spectral composition of the post burst emission (right pannel) generally fits 
to power law with indicies in the 1.3--2.5 range.

\begin{figure} [t!]
\plottwo{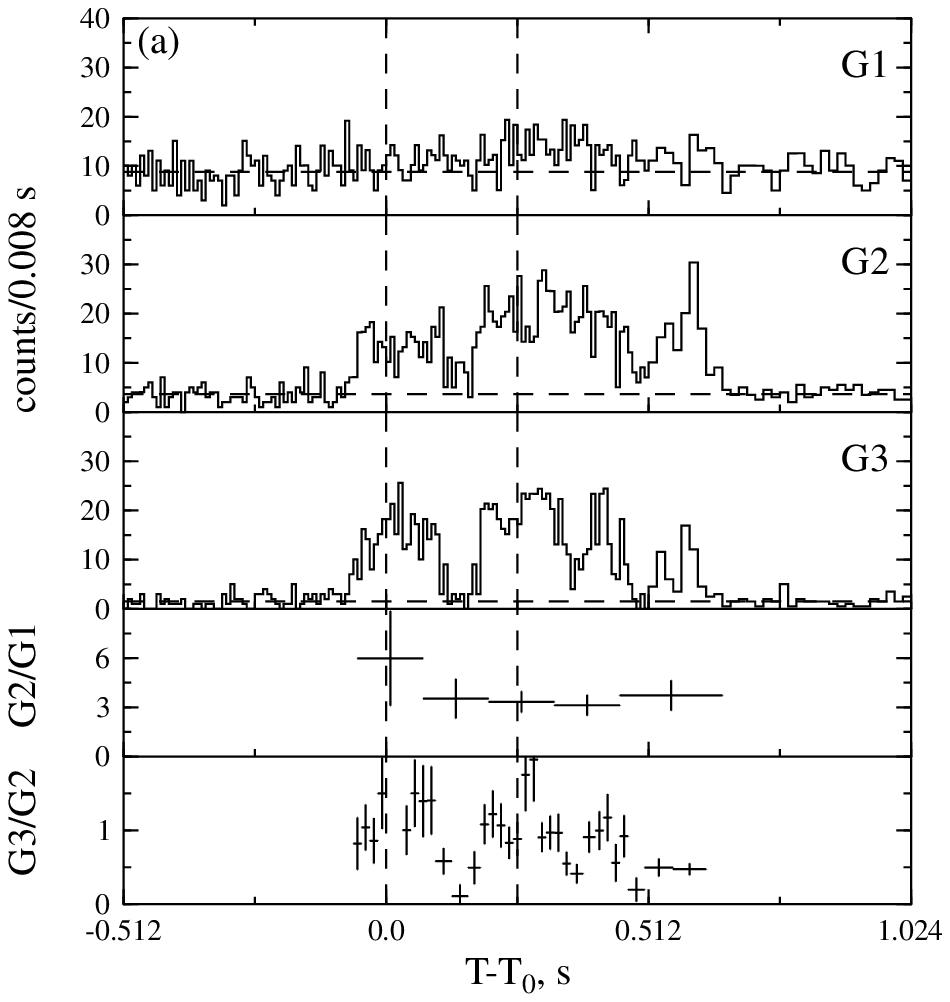}{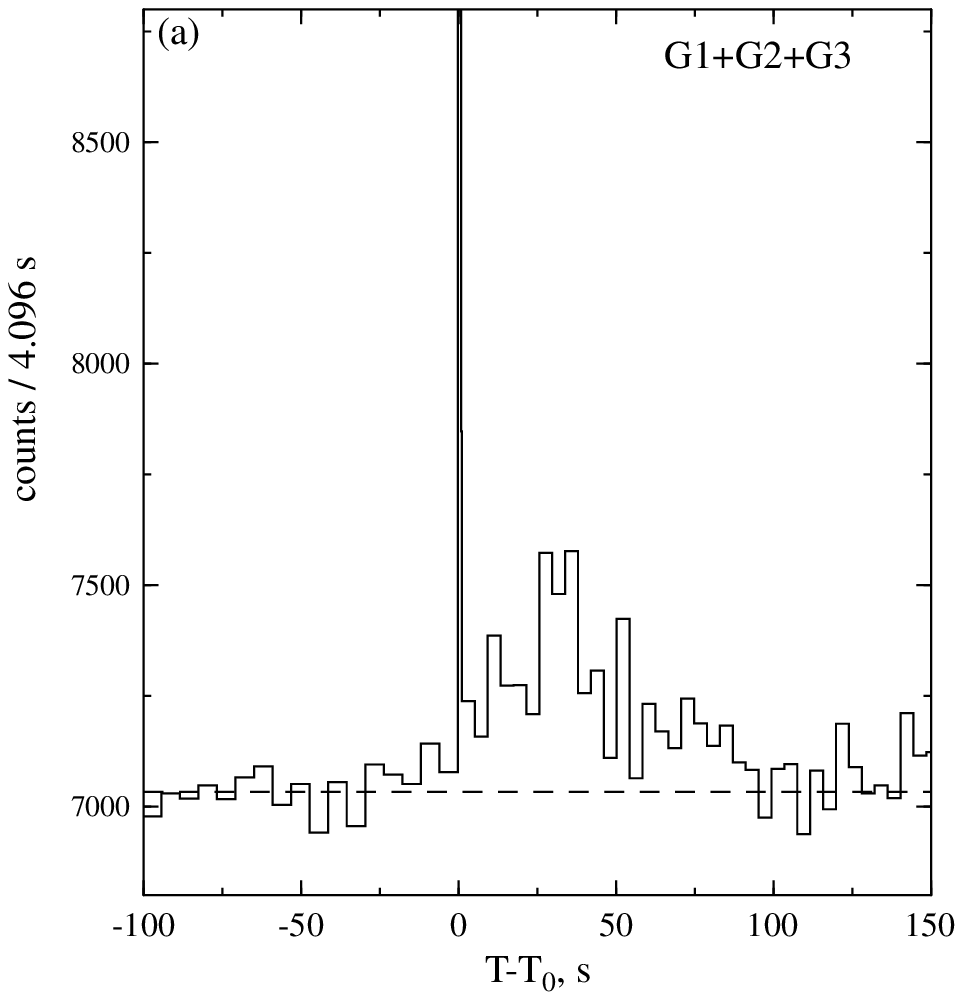}
\caption{Light curves of GRB990712a (left pannel) and postburst emission (right pannel).}
\end{figure}
\begin{figure} [b!]
\plottwo{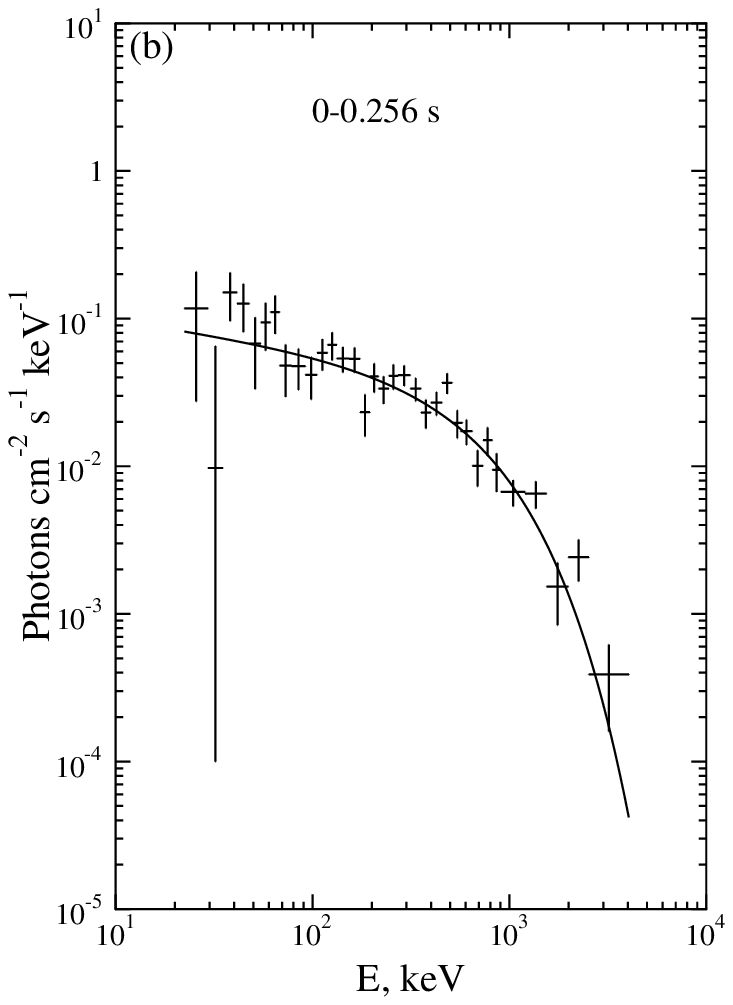}{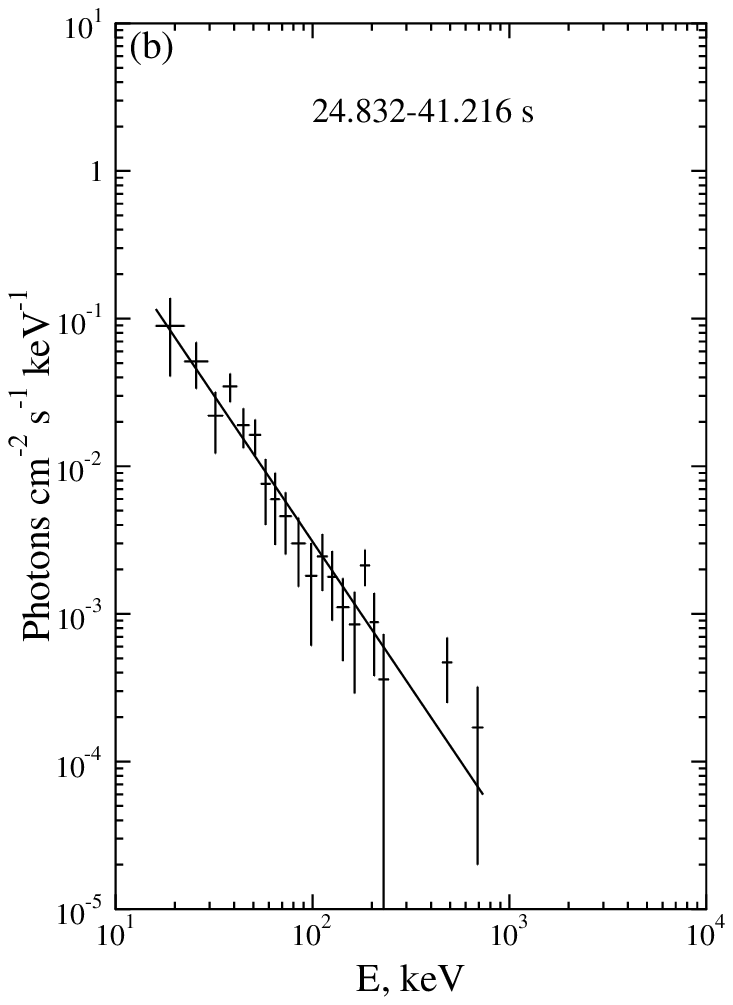}
\caption{Energy spectra of GRB990712a (left pannel) and postburst emission (right pannel).}
\end{figure}

\begin{figure} 
\plotone{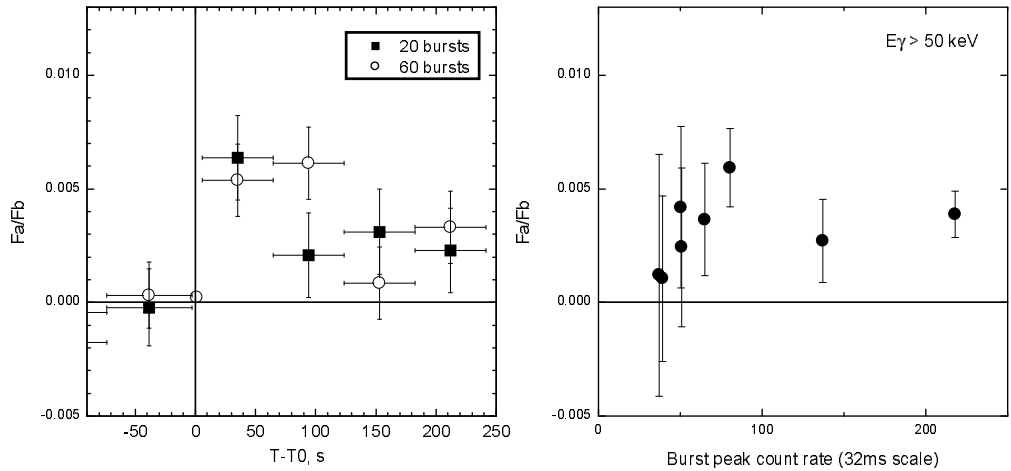}
\caption{Left: normalized afterglow flux for the samples of 20 and 60 background subtracted
and peak aligned bursts. Right: normalized afterglow flux (80 s) versus burst peak count rate (32 ms). 
Each point represents a 10 burst sample. }
\end{figure}

\section{Statistical Studies}

While other events don't exhibit such "visual" evidences, very stable 
radiation background conditions of the Konus-Wind experiment allow us to perform
search of post--burst emission on the summed up lightcurves of short GRBs not included in the
discussed set. 

For this analysis we used "background mode" 50--750 keV data sampled with bins of 2.944 s.
For the summed-up lightcurves of 100 events a post--burst emission at the level of 1.5$\pm$0.4
counts s$^{-1}$burst$^{-1}$ was found in 5--100 s interval after the GRB. Taking in consideration
an effective area of Konus-Wind detector this magnitude is comparable with (Connaughton, 2002).

Another interesting characteristic is a "normalized afterglow flux" Fa/Fb, i.e. 
a countrate divided by a "GRB bin" countrate. 
In Figure 3 (left pannel) Fa/Fb is plotted versus time after the trigger for the sums of 20 and 60 strongest events. 
On the right pannel this characteristic is plotted versus peak burst count rate for 90 events gruped by 10.
One can see that the magnitude of relative afterlow emission remains on the comparable level
for groups of different intensity GRBs, being limited only by statistcs for weaker events.
It leads us to a conclusion that an afterglow is a more common feature of short GRBs. 

\section{Concluding Remarks}

This work was supported by Russian Aviation and Space Agency Contract, 
and RFBR grant N 01-02-17808.

\end{document}